\documentclass[epj]{svjour}
\usepackage{graphicx,amsmath,amssymb}
\begin{document}
\title{Dipolar-controlled spin tunneling and relaxation in molecular magnets}
\author{D. A. Garanin 
}                     
%
%
\institute{ Department of Physics and Astronomy, Lehman College, City University of New York, \\
250 Bedford Park Boulevard West, Bronx, New York 10468, USA}
\date{Received: 12 October 2011}
%
\abstract{ Spin tunneling in molecular magnets controlled by dipole-dipole interactions
(DDI) in the disordered state has been considered numerically on the
basis of the microscopic model using the quantum mean-field approximation.
In the actual case of a strong DDI, coherence of spin tunneling is completely lost
and there is a slow relaxation of magnetization, described by $t^{3/4}$
at short times. Fast precessing nuclear spins, included in the model microscopically,
only moderately speed up the relaxation.
} 
\maketitle

\section{Introduction}

Molecular magnets (MM) built of molecules with a large effective spin,
such as $S=10$ in Mn$_{12}$ and Fe$_{8}$, attract attention by
their bistability resulting from a large uniaxial anisotropy $D$
\cite{sesgatcannov93nat,baretal96epl} that creates the energy barrier
$DS^{2}\simeq67$K for spin rotation (for a review, see \cite{gatcanparses94sci,sesgat03,gatsesvil06book}).
The most spectacular finding on molecular magnets is resonance spin
tunneling \cite{frisartejzio96prl,heretal96epl,thoetal96nat} that
occurs when spin energy levels on the different sides of the barrier
match each other. Another striking phenomenon is propagating fronts
of magnetic burning or deflagration \cite{suzetal05prl,heretal05prl},
a self-supporting process driven by the energy release that controls
thermally activated overbarrier relaxation \cite{garchu07prb}. As
further development, fronts of spin tunneling controlled by the dipolar
field \cite{garchu09prl,gar09prb} and a combined quantum-thermal
theory of magnetic deflagration \cite{garjaa10prbrc} have been proposed.

Magnetic molecules in MMs form a crystal lattice (bo\-dy-centered tetragonal
for Mn$_{12}$ Ac). As magnetic cores of the molecules are shielded
by organic ligands, there is no exchange interaction between the molecules
in the crystal, thus the dipole-dipole interaction (DDI) is dominating.
There is an evidence of dipolar ordering below 1 K in Mn$_{12}$ Ac and
other MMs \cite{luietal05prl,evaetal04prl,bowenetal10prb}, dynamics
of which is due to spin tunneling. Relaxation of spin states in molecular
magnets due to interaction with the environment should be a collective
process, as many molecular spins are interacting with the same phonon
or photon modes. Theories of photon and phonon superradiance in MMs
have been proposed in Refs. \cite{chugar02prl,chugar04prl}.

In the limit of low temperatures, processes of thermal activation
die out and the only way for the molecular spin to cross the barrier
is spin tunneling. Coherence of spin tunneling between the two degenerate
ground states in molecular magnets is difficult to observe because
the energy bias due to DDI is much greater than the tunnel splitting
of spin states $\Delta$. As a result, most of molecular spins are
strongly biased and cannot tunnel. Furthermore, tunneling of a spin
changes the dipolar field on other spins, prohibiting or allowing
them to tunnel. Thus low-temperature spin relaxation in molecular
magnets is a complicated collective process that results in a non-exponential
slow relaxation. The latter was initially observed as magnetization
relaxation from the saturated state in Fe$_{8}$ \cite{sanetal97prl}
and Mn$_{12}$Ac \cite{thocanbar99prl} that followed the $\sqrt{t}$
law. Theoretically this behavior had been explained by spreading of
the initially uniform dipolar field in samples of ellipsoidal shape
in the course of relaxation \cite{prosta98prl,cucforretadavil99epjb,chu00prl}.
This spreading gradually drives more and more spins off resonance
and slows down the relaxation. The situation is much less clear in
the realistic case of MM crystals of arbitrary shape where the dipolar
field is non-uniform from the beginning.

Tunneling of most of the spins being hampered by a large dipolar bias
leads to the quest for a mechanism that could accelerate spin transitions.
It was suggested that nuclear spins provide a fast fluctuating bias
on magnetic molecules that is much larger than $\Delta$ and brings
a larger number of molecules on resonance, allowing them to tunnel
\cite{prosta98prl}. However, combining formulas of Ref. \cite{prosta98prl}
leads to the final result for the relaxation rate that does not depend of nuclear spins \cite{chu00prl}.
The controversy penetrated experimental literature as well, where the formula
for the $\sqrt{t}$ relaxation rate (see, e.g., Eq. (2) of Ref. \cite{weretal99prl})
\begin{equation}
\Gamma_{1/2}\sim \Delta^{2} P(\xi_H)/\hbar,
\label{GammaSqrtP}\end{equation}
is used, $P(\xi_H)$ being the distribution of the energy bias on magnetic molecules.
As the latter is mainly due to the DDI, there is no place for nuclear spins in this formula.
With no external bias one has $P(\xi_H)\sim 1/E_D$ and \cite{chu00prl}
\begin{equation}
\Gamma_{1/2}\sim\Delta^{2}/(\hbar E_{D}),
\label{GammaSqrt}\end{equation}
where $E_{D}$ is the dipolar energy ($E_{D}/k_{B}$ is 0.067 K for Mn$_{12}$
Ac and 0.126 K for Fe$_8$).
With $\Delta/k_{B}\simeq 10^{-7}$ K for Fe$_8$ at zero field, the factor suppressing spin tunneling is very large:
$E_{D}/\Delta\sim 10^6$.

Further experimens showed that the $\sqrt{t}$ relaxation also follows
a small abrupt change of the bias field in the \emph{disordered} state
of MMs \cite{weretal99prl,weretal00prl}. Here it is difficult to
propose an analytical approach, and numerical methods had been applied.
Refs. \cite{feralo03prl,feralo04prb} state that the $\sqrt{t}$ relaxation
law is not universal, and for the body- and face-centered lattices
the relaxation exponent is 0.7 rather than 1/2.
These results were criticized in Refs. \cite{tupsta04prlcomm,tupstapro04prb,tupstapro05prbcomm}
as finite-size effect and effect of fitting the data over a too long
time interval, and $\sqrt{t}$ relaxation law was obtained for all
lattices at times short enough. However, Refs. \cite{feralo05prbcomm,feralo05prb}
refute the criticism, insisting on the non-universal behavior with
the relaxation exponent of up to 0.73 for the face-centered lattice.
At larger times, the magnetization relaxation follows a stretched
exponential \cite{feralo05prb}.

Recently dipolar-controlled relaxation together with dipolar ordering
has been observed in dynamic susceptibility experiments on an Er-based MM \cite{luisetal10prbrc}.

Most of published numerical work use Monte Carlo simulations based on the
``tunneling window'' concept, according to Ref. \cite{cucforretadavil99epjb}.
Spins are being checked one after the other, and if the bias on a spin
is within the preset tunneling window, the spin is allowed to flip.
The ensuing change of the dipolar fields is taken into account
when checking next spins. The tunneling window in these simulations
had been set to the amplitude of the fluctuating bias due to nuclear
spins that is much larger than $\Delta$. Although this approach captures
the essential physics of the DDI and leads to the results qualitatively
similar to what is seen in experiments, it is oversimplified and
postulates the role of nuclear spins instead of describing their
action dynamically. The ensuing relaxation rates are roughly proportional
to the tunneling window due to nuclear spins, that contradicts Eq.
(\ref{GammaSqrt}).
On the other hand, the absolute value of the relaxation rate cannot be found by Monte Carlo simulations
because Monte Carlo steps cannot be quantified in terms of real time without additional work.
The recent work \cite{vijgar11-arXiv} studying
low-temperature relaxation in molecular magnets out of saturation
with Monte Carlo and then rate equations, takes nuclear spins into
account in a different, although also phenomenological way. Ref. \cite{miysai01}
considers a one-molecule model with a stochastic random-walk-type bias, also obtaining
the $\sqrt{t}$ law.
Ref. \cite{sinpro03prb} considers the influence of nuclear spins on tunneling in the absence of DDI.

Theoretical methods mentioned above replace the original quantum-mechanical model by a more tractable sto\-chastic model.
Clearly, the ultimate solution of the problem should use the many-body Schr\"odinger equation.
This approach has been developed for a ``cental" system consisting of one spin or several interacting spins,
 coupled to a sea of other spins, e.g., nuclear spins. The most efficient numerical method is based on the  Chebyshev expansion
\cite{dobrae03pre,dobraekathar04HAIT}.
Since the number of coefficients in the expansion of the wave function of the system over a basis increases exponentially
with the system size, calculations are feasible for a number of spins up to 20 and require a supercomputer.
Decoherence and approach to equilibrium could be established with this method for small systems
\cite{sindob04prb,yuakatrae09jpsj,jinetal10jpsj}.
However, solution of the full Schr\"odinger equation for the system under consideration is problematic because the
dipole-dipole interaction is long-ranged and the size of the system has to be much larger.

The aim of the present work is to provide a microscopic description
of the dipolar-controlled tunneling relaxation in molecular magnets
on the basis of the real dynamics of molecular and nuclear spins,
and to clarify the long-standing controversy of the role of the latter.
The method consists in numerically solving the system of equations
of motion for molecular spins at equilibrium in the quantum mean-field
(Hartree) approximation and measuring the time-dependent auto-correlation function.
The latter with the help of the linear-response theory and other relations
can be proven to be proportional to the magnetization relaxation function
measured in experiments. This approach is presented in Sec. \ref{Sec-model}.
The numerical procedure is  described in Sec. \ref{Sec-numerical-solution}, while
Sec. \ref{Sec-numerical-results} contains numerical results for the Mn$_{12}$
Ac and simple cubic lattices in the absence of an external bias.
In Sec. \ref{Sec-bias} a stronger-than-expected influence of an external bias on spin relaxation is shown.
Section \ref{Sec-nuclear} introduces nuclear spins microscopically
via time-dependent bias fields acting on the molecular spins.
The results show a moderate speed-up of
the relaxation due to nuclear spins in the case of their fast precession.
In the concluding section, the
results obtained and their possible generalizations are discussed.

\section{The model and equations of motion}
\label{Sec-model}

At low temperatures, only the lowest doublet of the Hamiltonian of
a magnetic molecule is populated, so that one can use the two-state
model described by the pseudospin 1/2
\begin{equation}
\hat{H}=-\frac{\Delta}{2}\sum_{i}\sigma_{ix}-\frac{W_0}{2}\sum_{i}\sigma_{iz}-\frac{1}{2}\sum_{ij}V_{ij}\sigma_{iz}\sigma_{jz}.
\label{Ham}\end{equation}
Here $\boldsymbol{\sigma}$ is the Pauli matrix, $\Delta$ is tunnel splitting, $W_0=2Sg\mu_B B_z$ is the external bias
and $V_{ij}$ is the
dipole-dipole interaction\begin{equation}
V_{ij}=E_{D}\phi_{ij},\qquad\phi_{ij}=v_{0}\frac{3\cos\theta_{ij}-1}{r_{ij}^{3}},\label{DDI}\end{equation}
where \begin{equation}
E_{D}=(Sg\mu_{B})^{2}/v_{0}\label{ED}\end{equation}
is the dipolar energy ($E_{D}/k_{B}=0.067$K for Mn$_{12}$ Ac), $v_{0}$
is the unit-call volume, $r_{ij}$ is the distance between the molecules
at sites $i$ and $j$, and $\theta_{ij}$ is the angle between $\mathbf{r}_{ij}$
and the easy axis $z$. Eq. (\ref{Ham}) is a kind of transverse Ising
model with a long-range interaction.

The Heisenberg equation of motion for spin operators $\boldsymbol{\sigma}_{i}$
has the form\begin{equation}
\dot{\boldsymbol{\sigma}}_{i}(t)=\boldsymbol{\sigma}_{i}(t)\times\boldsymbol{\Omega}_{i},
\label{sigmaHeisEq}\end{equation}
where
\begin{equation}
\boldsymbol{\Omega}_{i}=\frac{1}{\hbar}\left(\Delta\mathbf{e}_{x}+W_{i}\mathbf{e}_{z}\right),\qquad
W_{i}=W_0+2\sum_{j}V_{ij}\sigma_{jz}.
\label{Omega}\end{equation}
One can see that Eq. (\ref{sigmaHeisEq}) describes precession of
the spin $\boldsymbol{\sigma}_{i}$ in an effective field. However,
going over from the Heisenberg operators to their quantum-mechanical
averages that we are interested in is notrivial because, in general,
different spin operators are entangled and their quantum averages
do not factorize, $\left\langle \sigma_{i\alpha}\sigma_{j\beta}\right\rangle \neq\left\langle \sigma_{i\alpha}\right\rangle \left\langle \sigma_{j\beta}\right\rangle $.
In the absence of factorization, one has to resort to the full Schr\"odin\-ger
equation for a many-spin system that can be solved only for a small
number of spins \cite{dobrae03pre,dobraekathar04HAIT,sindob04prb,yuakatrae09jpsj,jinetal10jpsj}.
To make the problem tractable numerically for a larger number of spins, one can
apply the quantum mean-field or Hartree approximation for spins
on different sites,\begin{equation}
\left\langle \sigma_{i\alpha}\sigma_{j\beta}\right\rangle \Rightarrow\left\langle \sigma_{i\alpha}\right\rangle \left\langle \sigma_{j\beta}\right\rangle ,\qquad i\neq j.\label{Decoupling}\end{equation}
This approximation means that each spin is exposed to the quantum-mechanically
averaged field from its neighbors, that should be justified when the
number of interacting neighbors is big. For the long-range dipole-dipole
interaction, the latter should be the case. After decoupling of quantum
correlations, the equation of motion for the averages of spin components
has the same form as Eq. (\ref{sigmaHeisEq}). Thus for the ease of
notation we replace $\left\langle \boldsymbol{\sigma_i}\right\rangle \Rightarrow\boldsymbol{\sigma_i}$
and obtain exactly Eq. (\ref{sigmaHeisEq}), now for precessing classical-like
vectors.

In the temperature range well above the dipolar ordering temperature
$\sim$1 K ($T=\infty$ at the scale of the DDI), molecular spins
are completely disordered and spins of different molecules are uncorrelated.
This is an additional justification for the decoupling in Eq. (\ref{Decoupling}).
The correlation function needed to calculate the spin relaxation
function thus reduces to the $zz$ component of the autocorrelation function
\begin{equation}
S_{i\alpha\beta}(t)\equiv\frac{1}{2}\left\langle \sigma_{i\alpha}(t)\sigma_{i\beta}(0)+\sigma_{i\beta}(0)\sigma_{i\alpha}(t)\right\rangle
\label{CFzz}\end{equation}
($\alpha,\beta=x,y,z$) that has to be averaged over all $N$ spins,
\begin{equation}
S_{zz}(t)\equiv\frac{1}{N}\sum_{i}S_{izz}(t).
\label{CFAveraged}\end{equation}
The autocorrelation function cannot be decoupled because this would
violate the properties of spin operators. One can
use the identities\begin{equation}
\sigma_{z}\sigma_{z}=\sigma_{x}\sigma_{x}=\sigma_{y}\sigma_{y}=1,\label{sigmaAC1}\end{equation}
and\begin{equation}
\sigma_{z}\sigma_{x}+\sigma_{x}\sigma_{z}=\sigma_{z}\sigma_{y}+\sigma_{y}\sigma_{z}=0\label{sigmaAC2}\end{equation}
to obtain the initial conditions for the autocorrelation function of Eq. (\ref{CFzz})
\begin{equation}
S_{i\alpha\beta}(0)=\begin{cases}
1, & \alpha=\beta\\
0, & \alpha\neq\beta.\end{cases}\label{CFIniCond}\end{equation}
The equation of motion for the autocorrelation function can be obtained
by differentiating over time, using the Heisenberg equation of motion,
Eq. (\ref{sigmaHeisEq}), and decoupling $\mathbf{\boldsymbol{\Omega}}_{i}$
since the latter is due to spins on other sites. This yields the equation
\begin{equation}
\frac{d}{dt}\mathbf{S}_{iz}(t)=\mathbf{S}_{iz}(t)\times\boldsymbol{\Omega}_{i},\label{CFEq}\end{equation}
where \begin{equation}
\mathbf{S}_{iz}(t)=\frac{1}{2}\left\langle \boldsymbol{\sigma}_{i}(t)\sigma_{iz}(0)+\sigma_{iz}(0)\boldsymbol{\sigma}_{i}(t)\right\rangle.
\label{CFVec}\end{equation}
Eqs. (\ref{sigmaHeisEq})
and (\ref{CFEq}) are coupled and have to be solved together. The
initial condition for Eq. (\ref{CFEq}) is $\mathbf{S}_{iz}(0)=\mathbf{e}_{z}$.
For Eq. (\ref{sigmaHeisEq}), we use the initial condition describing
$\boldsymbol{\sigma}_{i}(0)$ pointing in random directions. Note
that the autocorrelation function defined as $\sigma_{iz}(t)\sigma_{iz}(0)$
is wrong for MMs because it describes a system of classical rather
than of quantum spins. This model is of its own interest and it will be briefly addressed in Sec. \ref{Sec-classical}.

With the help of the linear response theory and other relations (see
Appendix) one can find the spin relaxation function measured in the experiment.
The latter turns out to be just $S_{zz}(t)$ of Eq. (\ref{CFAveraged}).
In particular,
if a small bias field $B_{z}$ is abruptly removed, the ensuing relaxation
is described by a remarcably simple formula\begin{equation}
\sigma_{z}(t)=\sigma_{z,\mathrm{eq}}S_{zz}(t),\label{SpinRelax:}\end{equation}
where $\sigma_{z,\mathrm{eq}}=2Sg\mu_{B}B_{z}/(k_{B}T)$ is the equilibrium
spin average in the high-temperature limit. Note that only $\sigma_{z,\mathrm{eq}}$
contains the temperature, whereas $S_{zz}(t)$ can be calculated at
$T=\infty$. In other cases, such as a small field $B_{z}$ abruptly
applied to an initially unbiased MM at equilibrium, one obtains similar
formulas.

\section{Numerical solution}
\label{Sec-numerical-solution}

Numerical solution of Eqs. (\ref{sigmaHeisEq}) and (\ref{CFEq})
is a serious task because there are six dynamic variables per lattice
site and the DDI is long-ranged.
The latter can be dealt with by methods based on the fast Fourier transform (see, e.g., Ref. \cite{hinnow00jmmm}).
Fortunately, the autocorrelation
function self-averages over random initial orientations of spins.
Thus for a very large number of molecular spins the solution is practically
the same for any realization of the initial conditions.
However, when the number of spins
is not extremely large, averaging over several runs with different realizations of initial conditions
is needed to make the time dependence of the autocorrelation function smoother.

In most of the calculations, the body-centered tetragonal lattice
of Mn$_{12}$Ac has been used.
Some calculations have been performed for a simple cubic lattice to check universality of the relaxation exponent.
Numerically it is impossible to deal with the realistic case of very large $E_D/\Delta$ because the relaxation becomes too slow and computation time becomes too long.
However, the large-$E_D$ behavior sets in already for $E_D/\Delta\geq 3$, and the results for $3\geq E_D/\Delta\geq 100$ can be scaled to obtain the result for any large $E_D/\Delta$.

The sizes of crystals are expressed in units of $a$ for the body-centered tetragonal
lattice of Mn$_{12}$ Ac with lattice parameters $a=b>c$.
The ``external" sublattice contains $N_a N_b N_c$ spins while the
``internal" sublattice contains $(N_a-1)(N_b-1)(N_c-1)$ spins, where $N_a=L_a+1$, $N_b=L_b+1$, and
$N_c={\rm Round}(L_c/\eta)+1$ with $\eta=c/a$.
The total number of spins is $N=N_a N_b N_c+(N_a-1)(N_b-1)(N_c-1)$.
In most calculations, $N$ was above 20000 that is comparable with the system sizes in Monte Carlo simulations of
Ref. \cite{cucforretadavil99epjb} and subsequent works.

The rate of spin tunneling (i.e., precession of the spin vector around the $x$ axis) is determined by $\Delta$.
Most of the time, however, the bias of the spin is very large, $W_i \gg \Delta$, and the spin vector is fast precessing around the $z$ axis.
This fast precession does not contribute to spin tunneling and is actually irrelevant.
However, it necessitates a very small integration step in numerically solving the equations of motion, to achieve a sufficient accuracy.
For numerical efficiency, it is absolutely crucial to introduce a cutoff of the bias $W_i$ that allows to increase the integration step.
In these calculations, a cut-off at $W_{\rm max} = 5\Delta$ was introduced by replacing in the equations of motion
$W_i\Rightarrow W_{i,\rm eff}=W_i/Q_i$ and $\Delta\Rightarrow \Delta_{i,\rm eff}=\Delta/Q_i$,
where $Q_i=\sqrt{1+(W_i/W_{\rm max})^c}$ and $c=6$.

Numerical calculations have been performed using Wolfram Mathematica 8.0.4.0 that supports vectorization and makes use of Intel's
math kernel library (MKL) for number crunching.
The latter employs threading for standard tasks such as solving vectorized systems
of ordinary differential equations (ODE), that results in a significant speed-up on multi-processor machines for a large system size.
The most efficient algorithm for solving our ODE's proved to be Runge-Kutta 4th order with a fixed step size.
The latter was chosen so that the average deviation of the spin length from 1 at the end of the calculation does not exceed 1\%.
The computers used were (i) Mac Pro with two 2.4 GHz quad-core Intel Xeon processors and 16 GB memory and
(ii) Lenovo Y570 laptop with Intel i7-2670QM 2.2 GHz processor and 8 GB memory.
With the turbo boost processor frequency of 3.1 GHz, the Lenovo laptop is faster than Mac Pro and has a higher Mathematica benchmark,
1.1 vs 0.7.
However, with the double number of processor cores, Mac Pro has an edge in the overall speed of our computations.
The computation time was about two days for a typical relaxation curve.

\section{Numerical results}
\label{Sec-numerical-results}

The dynamics of the system in the absence of the external bias and nuclear spins is in accord with the tunneling-window
concept. For any given spin, most of the time $W_i\gg\Delta$ in Eq.
(\ref{Omega}), so that the spin is merely precessing around the $z$
axis and no tunneling occurs. Only when $W_i\lesssim\Delta$ (dipolar
bias within the tunneling window) the spin rotates around the $x$
axis and $\sigma_{z}$ changes. For $E_{D}\gg\Delta$ tunneling
events are rare and spin relaxation is very slow, that results in
 long computation times. In this limit, large system sizes are
mandatory, because for the number of spins not large enough it can happen
that no spin is within the tunneling window and the relaxation
gets stuck forever.

\begin{figure}
\includegraphics[angle=-90,width=8cm]{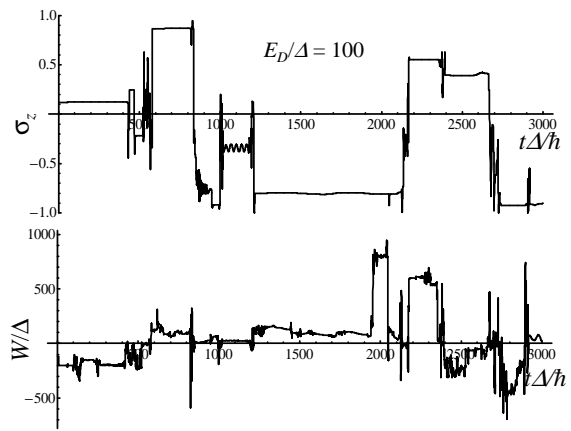}
\caption{Typical time dependence of spin average and the energy bias of a magnetic molecule.
Change of $\sigma_z$ occurs when $W$ goes through zero.
\label{Fig-sigz-W-Mn_12-box_Labc=00003D4_ED=00003D100} }
\end{figure}

Typical time dependence of
the spin polarization $\sigma_{z}$ and the energy bias $W$ of a
magnetic molecule is shown in Fig. \ref{Fig-sigz-W-Mn_12-box_Labc=00003D4_ED=00003D100}.
One can see that $\sigma_{z}$ changes only when
$W$ approaches zero and both time dependences are stationary quasi-random
processes. Note that $E_{D}$ is a measure of the dipolar field produced
by one spin on its neighbors. As dipolar fields from different spins
add up, the total bias $W$ in Fig. \ref{Fig-sigz-W-Mn_12-box_Labc=00003D4_ED=00003D100} exceeds $E_{D}$ by a factor of up
to 10.
\begin{figure}
\includegraphics[angle=-90,width=8cm]{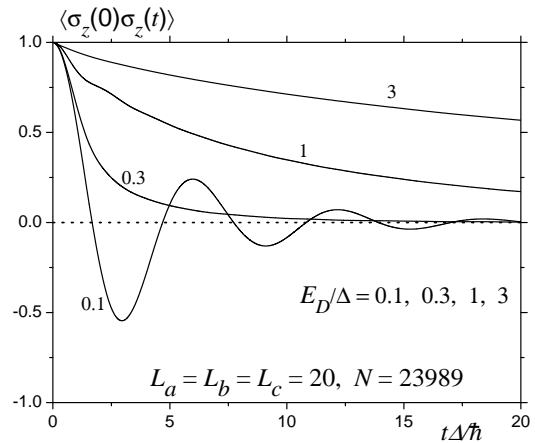}\caption{Crossover of the spin auto-correlation function from
weak to strong DDI that suppresses coherence of spin tunneling.\label{Fig-CF-Mn_12-box_Labc=00003D6-small_ED} }
\end{figure}
\begin{figure}
\includegraphics[angle=-90,width=8cm]{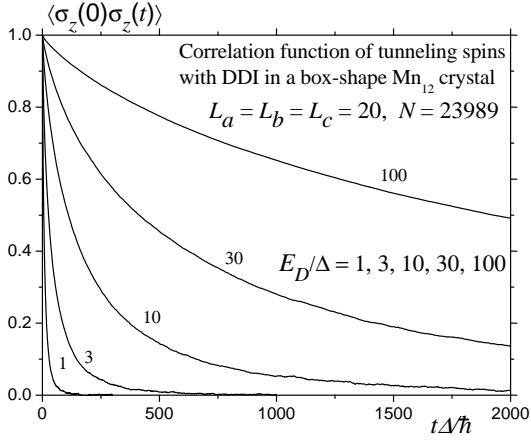}
\caption{Spin autocorrelation function for strong DDI is decaying very slowly.
\label{Fig-CF-Mn_12-box_Labc=00003D6}}
\end{figure}

Let us consider now the results for the spin autocorrelation function. The results for $E_{D}/\Delta\sim1$
are shown in Fig. \ref{Fig-CF-Mn_12-box_Labc=00003D6-small_ED}. For
a weak dipolar interaction $E_{D}$ (or large tunnel splitting $\Delta$)
there are damped oscillations at frequency $\Delta/\hbar$ showing
coherence, however. Although $\Delta$ can be increased by a strong
transverse field, $E_{D}/\Delta$ always remains a large parameter.
With increasing $E_{D}/\Delta$ in Fig. \ref{Fig-CF-Mn_12-box_Labc=00003D6-small_ED},
oscillations disappear and spin tunneling becomes incoherent.

Figure
\ref{Fig-CF-Mn_12-box_Labc=00003D6} shows that for a strong dipole-dipole
interaction spin relaxation becomes very slow.
\begin{figure}
\includegraphics[angle=-90,width=8cm]{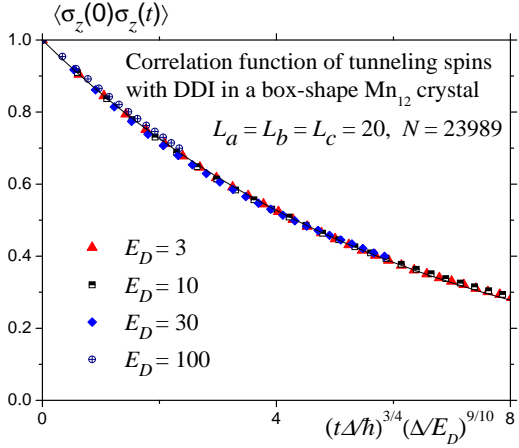}
\caption{Spin autocorrelation function in the scaled form.
Solid line is the stretched exponential of Eq. (\ref{StretchedExp}).
\label{Fig-CF-Mn_12-box_Labc=00003D6-scaled}}
\end{figure}
The results of Fig. \ref{Fig-CF-Mn_12-box_Labc=00003D6} for strong
dipolar field are represented in the scaled form in Fig. \ref{Fig-CF-Mn_12-box_Labc=00003D6-scaled}
that shows a $t^{3/4}$ law for the magnetization relaxation at short times.
The
results can be fitted to the stretched exponential
\begin{equation}
S_{zz}(t)=e^{-0.16\left(t\Gamma_{3/4}\right)^{3/4}},
\qquad\Gamma_{3/4}=\frac{\Delta}{\hbar}\left(\frac{\Delta}{E_D}\right)^{6/5}.
\label{StretchedExp}\end{equation}
 Both the exponent in the spin relaxation function and the
relaxation rate differ from the previously obtained result of
Eq. (\ref{GammaSqrt}).
Although numerical calculations have been performed for Mn$_{12}$ Ac, one can expect the same result,
up to a numerical coefficient, also for Fe$_8$.
Using parameters below Eq. (\ref{GammaSqrt}) for Fe$_8$, one obtains $\Gamma_{3/4} \simeq 10^{-4}$ s$^{-1}$.
This by an order of magnitude exceeds the $t^{1/2}$ relaxation rate $\Gamma_{1/2} \simeq 10^{-5}$ s$^{-1}$ measured in completely disordered
Fe$_8$ at zero external bias, Fig. 3 of Ref. \cite{weretal99prl}.
The relatively fast spin relaxation in the model with only dipolar interactions does not support the narrative of DDI
blocking spin tunneling so strongly that rapidly fluctuating nuclear spins are needed to open up a bigger tunneling window.
In fact, this conclusion already follows from Eq. (\ref{GammaSqrt}).

\begin{figure}
\includegraphics[angle=-90,width=8cm]{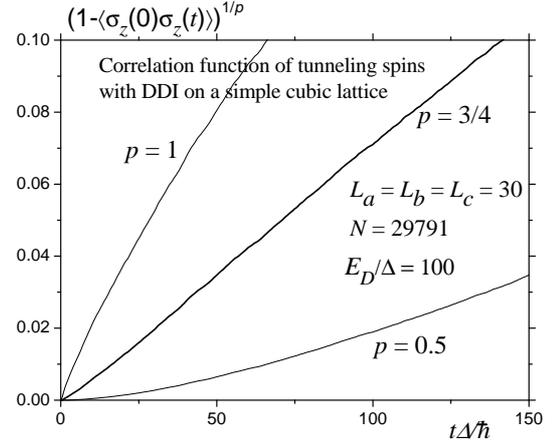}
\caption{Testing different exponents $p$ for the short-time quantum spin CF,
 $S_{zz}(t)\cong 1- a t^p$, on a simple cubic
lattice.\label{Fig-CF-sc-box_Labc=00003D8-power_p}}
\end{figure}
As another test for the short-time dependence of the spin autocorrelation
function, calculations for the simple cubic lattice have been done.
In Fig. \ref{Fig-CF-sc-box_Labc=00003D8-power_p}, this dependence
is shown in three forms that allows one to choose between three different
values for the time exponent. Clearly, $p=3/4$ comes out as a winner.
The same value of this exponent for the Mn$_{12}$ Ac body-cented
tetragonal lattice and the simple cubic lattice suggests universality
of spin relaxation.

\section{Influence of a static external bias}
\label{Sec-bias}

\begin{figure}
\includegraphics[angle=-90,width=8cm]{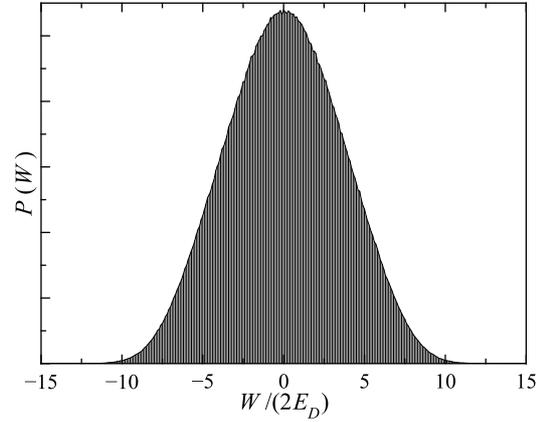}
\caption{Distribution of the dipolar bias on a magnetic molecule in Mn$_{12}$ Ac in a disordered state is nearly Gaussian with a small triangular distortion and extends up to $W\simeq 20E_D$.
\label{Fig-Dipolar_distribution}}
\end{figure}
\begin{figure}
\includegraphics[angle=-90,width=8cm]{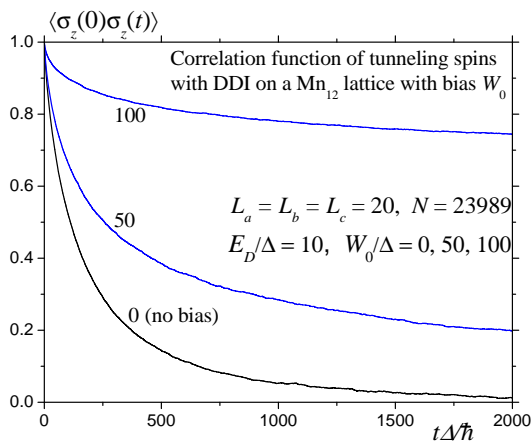}
\caption{Spin autocorrelation function of Mn$_{12}$ with a static external bias $W_0$.
Suppression of relaxation is much stronger than modelled by
the distribution of the dipolar bias in Fig. \ref{Fig-Dipolar_distribution}.
\label{Fig-CF-Mn_12-box_Labc=20-biased}}
\end{figure}

Static external bias $W_0$ in Eq. (\ref{Ham}) should slow down the relaxation because $W_0$ has to be compensated for by the dipolar bias,
that makes less spins being on resonance at any moment of time.
The expected dependence of the relaxation rate on $W_0$ is given by $P(W_0)$, the distribution of the dipolar bias that enters Eq. (\ref{GammaSqrtP}).
For a big crystal of Mn$_{12}$ Ac in a fully disordered state $P(W)$ shown in Fig. \ref{Fig-Dipolar_distribution} is nearly Gaussian
with a small triangular distortion and it extends up to $W\simeq 20E_D$.
Here the fully disordered state has been modelled by spin vectors $\boldsymbol{\sigma}_{i}(0)$ pointing in random directions, as in the dynamical calculations.

Surprisingly, the results of numerical calculations in Fig. \ref{Fig-CF-Mn_12-box_Labc=20-biased}
show a much stronger suppression of the spin relaxation by the external bias than conjectured above.
Indeed, for the parameters in Fig. \ref{Fig-CF-Mn_12-box_Labc=20-biased}, $W_0/\Delta=100$ corresponds to $W_0/(2E_D)=5$.
According to Eq. (\ref{GammaSqrtP}) and Fig. \ref{Fig-Dipolar_distribution}, the relaxation rate should drop by a factor of two.
However, the actual slow-down of spin relaxation is so strong that it suggests that a significant part of spins are not relaxing at all,
at least at the scale of these computations.
For $W_0/\Delta=50$ (i.e., $W_0/(2E_D)=2.5$) one can expect only a small effect of the static bias.
However, the relaxation becomes very slow at long times.

A likely reason for this behavior is the following.
Significant dipolar bias in Mn$_{12}$ is due to regions of spins aligned in the same direction.
Thus spins with a bias of another sign must be spatially well separated from the first group of spins.
Because of this large spatial separation, tunneling of the spins in the first group does not change much the dipolar bias on the spins in the
second group.
As the result, the latter have to wait a very long time until their dipolar bias changes via slow spatial diffusion and they come on resonance.

Since the shape of the relaxation curves at nonzero external bias is unclear, it is difficult to extract the dependence of the relaxation rate on the bias.
This problem requires further investigation.

\section{``Classical" spin correlation function}
\label{Sec-classical}

As a cruder approximation, instead of Eq. (\ref{CFzz}) one could use the classical-like spin CF defined by
\begin{equation}
{\cal S}_{izz}(t)\equiv\sigma_{iz}(t)\sigma_{iz}(0),
\label{CCFzz}\end{equation}
where $\sigma_{iz}(t)$ follows from Eq. (\ref{sigmaHeisEq}).
Although this definition violates quantum properties of spins 1/2, it still captures most of the physics and is still
superior to the Monte Carlo method as it is directly follows from the original model.
Since the initial state is a random, the average of ${\cal S}_{izz}(0)$ is 1/3, and thus one must define the average
autocorrelation function as
\begin{equation}
{\cal S}_{zz}(t)\equiv\frac{3}{N}\sum_{i}{\cal S}_{izz}(t)
\label{CCFAveraged}\end{equation}
that satisfies ${\cal S}_{zz}(0)=1$ in the thermodynamic limit.

\begin{figure}
\includegraphics[angle=-90,width=8cm]{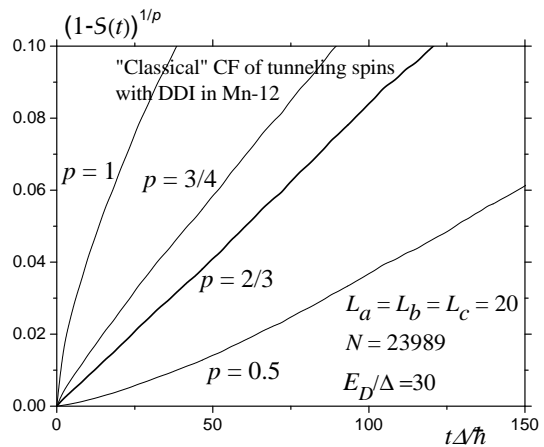}
\caption{Testing different exponents $p$ for the short-time ``classical" spin CF,
 ${\cal S}_{zz}(t)\cong 1- a t^p$, on the Mn$_{12}$
lattice.\label{Fig-CCF-sc-box_Labc=20-power_p}}
\end{figure}

As method of ``classical" CFs involves only 3 dynamic variables per lattice site, it is two times faster than the
method using quantum CFs. The short-time behavior of the ``classical" CF is shown in
Fig. \ref{Fig-CCF-sc-box_Labc=20-power_p}.
In contrast to Fig. \ref{Fig-CF-sc-box_Labc=00003D8-power_p}, the results can be best fitted with $p=2/3$
in ${\cal S}_{zz}(t)\cong 1- a t^p$, that differs from $p=3/4$ for the quantum spin CF above.

\section{Influence of nuclear spins}
\label{Sec-nuclear}

Magnetic atoms of some molecular magnets, such as Mn$_{12}$, possess
nuclear spins, while others such as the main isotope of Fe in Fe$_{8},$
do not. These nuclear spins are coupled to the electronic spins by
the hyperfine interaction that for them is stronger than any other
interaction. As a result, nuclear spins are precessing in the effective
field created by the electronic spins and produce a random but static
bias on the latter. This can essentially modify the Landau-Zener effect
\cite{garsch05prb,garnebsch08prb} but does not provide a fluctuating
bias that could extend the tunneling window.

On the other hand, all magnetic molecules contain a large number of
protons (in the hydrogen atoms of their ligands) whose nuclear spins
interact with electronic spins by the nuclear dipole-dipole interaction.
As in the case of the DDI, the only realistic way to consider the system's dynamics is to make
decoupling of quantum correlators similar to Eq. (\ref{Decoupling}).
After that nuclear spins can be considered as time-dependent bias on electronic spins.

Since protons have nuclear spin $I=1/2$, there are no quadrupolar terms in their Hamiltonian,
and their dynamics is precession in the magnetic field $B$,
\begin{equation}
\mathbf{\dot{I}}=(\mu_{p}/\hbar)\left[\mathbf{I\times B}\right],\qquad\mu_{p}=2.79\mu_{n},
\label{EqMotionProtons}\end{equation}
with the frequency $\omega_{p}=(\mu_{p}/\hbar)B$.
Here $\mu_p$ is the magnetic moment of the proton and $\mu_{n}$ is the nuclear magneton.
If no external field is applied, the dominating magnetic field is dipolar field from the electronic spins.
While the fully ordered state of Mn$_{12}$ Ac creates the dipolar field of 53 mT, disordered states of
Mn$_{12}$ Ac and Fe$_8$ create a somewhat smaller field.
With $B=16$ mT one obtains $\omega_{p}=1.3\times 10^6$ s$^{-1}$ that largely exceeds the tunneling frequency $\Delta/\hbar$.
For Fe$_8$ in zero field $\Delta/\hbar=1.3 \times 10^4$ s$^{-1}$, so that $\hbar\omega_p/\Delta \simeq 10^2$.
If a strong transverse field $B_\bot$ is applied to increase $\Delta$, precession of nuclear spins becomes regular with
the fixed frequency $\omega_{p}=(\mu_{p}/\hbar)B_\bot$.
For Fe$_8$, the ratio $\hbar\omega_p/\Delta$ initially strongly increases with $B_\bot$ but then decreases as $\Delta$ begins to increase
as a high power of $B_\bot$ \cite{gar91jpa} (see experimental data in Ref. \cite{werses99science}).
In all cases $\hbar\omega_p/\Delta$ is a large parameter.

To estimate the fluctuating bias from the protons, one can use the
data for the tunneling resonance spread of approximately $B_{p}\simeq0.5$
mT (data for Fe$_{8}$ from Ref. \cite{weretal00prl}).
As follows from the properties of the dipolar interaction, a part of this bias is static and thus irrelevant here,
whereas its another part is precessing with a frequency $\omega_p$.
The latter creates a time-dependent bias
\begin{equation}
W_{i}^{(p)}(t)=W_{p}\cos\left(\omega_{p}t+\varphi_{i}\right),\quad W_{p}=2Sg\mu_{B}B_{p}.
\label{Wprotons}\end{equation}
In contrast to the model used in  Ref. \cite{prosta98prl}, relevant dynamics of nuclear spins is precession, rather than their dephasing at rate $T_2^{-1}$.
Here, one could take into account distribution of $W_p$ values, as well as possible distribution of $\omega_p$.
However, this does not bring any qualitative change of the results because the dominating DDI already creates enough randomness
for a complete decoherence.
For this reason and the sake of simplicity, numerical calculations have been performed for the fixed ratio $W_p/(2E_D)=0.1$ corresponding to molecular magnets
and different fixed values of the ratio $\hbar\omega_p/\Delta$.
Note that the nuclear bias $W_p$ is by a factor of order 10$^2$ smaller than the dipolar bias, the distribution of which is shown in Fig. \ref{Fig-Dipolar_distribution}.

Let us check whether the bias created by precessing nuclear
spins of protons is fast, as required \cite{prosta98prl}, by considering
Landau-Zener transitions of electronic spins.
The typical value of
the LZ parameter
\begin{equation}
\varepsilon=\frac{\pi\Delta^{2}}{2\hbar v}
\label{epsilonLZ}\end{equation}
 where $v$ is the energy sweep rate. With $v=dW_{i}^{(p)}(t)/dt\sim W_{p}\omega_{p}$
it can be estimated as
\begin{equation}
\varepsilon\sim\frac{\pi\Delta^{2}}{2\hbar\omega_{p}W_{p}}.
\label{epsWpomp}\end{equation}
Since both $W_p/\Delta \gg 1$ and $\hbar\omega_p/\Delta \gg 1$, one has $\varepsilon \ll 1$ and the sweep is indeed fast.

\begin{figure}
\includegraphics[angle=-90,width=8cm]{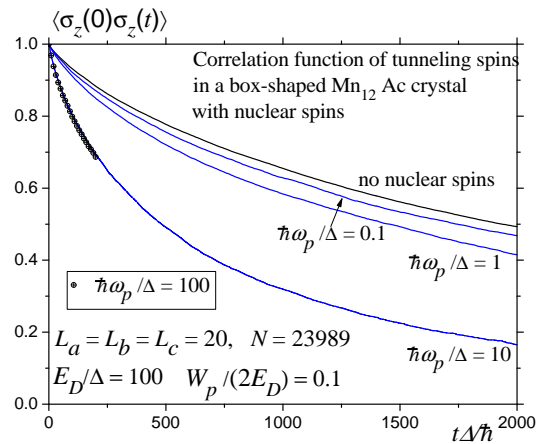}
\caption{Spin correlation function taking into account nuclear spins.
The latter speed up the relaxation by a factor of about 3 in the limit of rapid precession.
\label{Fig-CF-Mn_12-box_Labc=00003D4_ED=00003D100_protons}}

\end{figure}

Numerically, it is difficult to perform calculations for a very rapidly oscillating nuclear bias, Eq. (\ref{Wprotons}), with the same method as above
because it requires integration of equations of motion with a very small time step over a long time.
Whereas an appropriate semi-analytical method will be considered elsewhere, here we just perform numerical calculations for moderately high
values of $\hbar\omega_p/\Delta$ to see the effect of nuclear spins.
It can be expected that in the case of a slow nuclear precession, $\varepsilon\lesssim 1$,
the effect of nuclear spins is not strong.
One cannot speak of widening of the tunneling window because some spins are getting a chance to tunnel because of $W_p$ while other spins
lose their chance.
Thus, the tunneling chance is merely redistributed between spins but no additional tunneling chance is created.

On the other hand, one can speak of forced spin transitions via the Landau-Zener (LZ) effect.
It has been shown that interactions, including the DDI, severely suppress LZ transitions \cite{garsch05prb}.
For a fast sweep, however, interactions do not play a big role, and the standard LZ effect with a small transition probability takes place.
In spite of a small transition probability at fast sweep, there are repeated level crossings for an oscillating bias,
and the effect of many crossings accumulate.
It was shown that a combination of a time-linear sweep and a rapidly oscillating sweep results in the same transition probability
as the pure linear sweep \cite{delbarcounpub}.
Thus one can expect a more substantial effect of nuclear spins on low-temperature relaxation in molecular magnets in the realistic case
of rapidly precessing nuclear spins.
This effect should saturate at high precession rates of nuclear spins.

The results of numerical calculations in Fig. \ref{Fig-CF-Mn_12-box_Labc=00003D4_ED=00003D100_protons}
confirm the theoretical conjectures made above.
In the case of
a slow precession $\hbar\omega_{p}/\Delta=0.1$ (thus $\varepsilon\sim 1$) the effect of nuclear spins is small.
For a fast precession $\hbar\omega_{p}/\Delta=10$ (thus $\varepsilon\sim 0.01$)
it is substantial, creating a speed-up of relaxation by a factor of about 3.
However, such a speed-up is not a profound effect that could be expected from the argument that
tunneling window $\Delta$ is replaced by that of $W_p\gg\Delta$.
In fact, this effect should be somewhat smaller because only a part of the nuclear bias is oscillating while another part is static.
On the other hand, spin relaxation in the absence of nuclear spins, Eq. (\ref{StretchedExp}), is already fast enough,
so that one really does not need an additional speed-up from nuclear spins to describe experiments.

It is difficult to further increase $\omega_{p}$ within the present computational scheme and time limitations.
However, $\varepsilon\sim 0.01$ for $\hbar\omega_{p}/\Delta=10$ means already a pretty fast sweep,
and the results can be expected to saturate in the fast-sweep limit.
Indeed, calculation for $\hbar\omega_{p}/\Delta=100$ with a 10 times smaller integration step and 10 times shorter time interval
(to preserve the computing time) shows the same results as for $\hbar\omega_{p}/\Delta=10$.

\section{Discussion}

Results for the dipolar-controlled low-temperature spin relaxation
in molecular magnets in the disordered state presented above have
been obtained with a direct method based on the actual quantum spin dynamics of
the system. An essential approximation made is decoupling of quantum
correlations at different sites, the quantum mean field or Hartree
approximation. Although the results of this approximation may differ
from the exact solution of the Schr\"odinger equation (in particular,
for the Landau-Zener effect in systems with DDI \cite{garsch05prb}),
one can expect that this approximation captures the essential physics
and its results are qualitatively correct. In any case, this approach
is more fundamental than previous Monte Carlo simulations and yields the relaxation
curves in terms of the real time rather than of Monte Carlo steps.

Most of the results have been obtained for Mn$_{12}$Ac lattice in the case of a pure DDI without nuclear spins.
For other molecular magnets such as Fe$_8$ the results should be qualitatively the same.
The results show that relaxation is due to rare tunneling events in
the actual case of random dipolar bias much greater than tunnel splitting.
The relaxation law is a stretched exponential in the range of not too long times.
 The exponent in the stretched
exponential extracted from fits to the data obtained is
$p=3/4$ for both Mn$_{12}$ Ac and simple cubic lattices, differing
from the experimental law $t^{1/2}$ \cite{weretal99prl,weretal00prl}
and Monte Carlo results of Refs. \cite{tupsta04prlcomm,tupstapro04prb,tupstapro05prbcomm}.
On the other hand, it is in a qualitative accord with Monte Carlo
results of Refs. \cite{feralo05prbcomm,feralo05prb}, $p=0.7-0.73$
for a face-centered lattice.

The effect of nuclear spins has been included on the basis of a simplified  microscopic
model and it has been shown that nuclear spins speed-up
the relaxation in the realistic case of their fast precession.
However, the observed speed-up is moderate and it does not support the idea of nuclear spins opening a huge
 tunneling window.

The deviation of the relaxation law obtained above from the experimentally
observed may be a consequence of either thermally assisted tunneling
or correlations accompanying dipolar ordering, or both. The problem
is that to get rid of the ordering effects, the temperature has to
be above 1 K. However, at such temperatures thermally-assisted tunneling
should be already strong. Although populations of all states except
of the ground-state doublet are still negligible, there is a competition
between different ways to cross the barrier, pure tunneling considered
here and thermally-assisted tunneling over higher doublets. Since
the tunnel splitting strongly increases with the energy, thermally-assisted
tunneling remains competitive down to the temperatures below 1 K \cite{chugar97prl,garchu97prb}.
Thus the theory developed above may have no applicability range, strictly
speaking. Still it makes sense as the most basic microscopic theory
of dipolar controlled spin tunneling. It should be noted that none
of the preceding theoretical works on low-temperature spin relaxation
dealt with these two effects, dipolar ordering and thermally assisted
tunneling. The only investigation of the interplay of the dipolar
ordering and dipolar controlled relaxation is that in the recent experimental
work on Er, Ref. \cite{luisetal10prbrc}.

The discussion above shows the tasks of future investigations. It
would be interesting to extend the method to higher temperatures into
the regime of thermally assisted tunneling, as well as to lower temperatures
into the region of dipolar ordering. The latter is challenging because
it requires preparation of the initial state having a particular dipolar
energy corresponding to a given temperature. As at finite temperatures
there are spin-spin correlations, one cannot use the autocorrelations
function. As a result, self-averaging does not occur and one has to
make averaging over many initial states, even for a system of a large
number of spins.

\section*{Acknowledgments}
This work has been supported by the NSF under Grant No. DMR-0703639.
The author thanks E. M. Chudnovsky
for useful discussions.

\section*{Appendix: Spin relaxation function via spin correlation function}

Let us define the linear response of the normalized pseudospin with
respect to the energy variable $\xi=Sg\mu_{B}B_{z}.$ The linear response
has the form \begin{equation}
\left\langle \sigma_{z}\right\rangle _{t}=\int_{-\infty}^{\infty}dt^{\prime}L(t-t^{\prime})\xi(t^{\prime}),\label{LinResp}\end{equation}
where, because of the causality, $L(\tau)=0$ for $\tau<0.$ We are
interested in the response to the step $\xi(t)=\xi_{0}\theta(-t)$
(field switched off at $t=0$) that reads \begin{equation}
\left\langle \sigma_{z}\right\rangle _{t}=\xi_{0}\int_{-\infty}^{0}dt^{\prime}L(t-t^{\prime}),\qquad t>0.\label{LinRespStep}\end{equation}
 It is convenient to express this response in terms of the dynamic
susceptibility. Response to the oscillating field $\xi(t)=\xi_{0}e^{-i\omega t}$
has the form \begin{equation}
\left\langle \sigma_{z}\right\rangle _{t}=\xi_{0}\int_{-\infty}^{\infty}dt^{\prime}L(t-t^{\prime})e^{-i\omega t'}=\xi_{0}e^{-i\omega t}\chi(\omega),\end{equation}
 where \begin{equation}
\chi(\omega)=\int_{-\infty}^{\infty}d\tau L(\tau)e^{i\omega\tau}\label{chiviaL}\end{equation}
 is the susceptibility with respect to $\xi.$ Inverting this formula,
one obtains \begin{equation}
L(\tau)=\int_{-\infty}^{\infty}\frac{d\omega}{2\pi}e^{-i\omega\tau-\epsilon\tau}\chi(\omega),\qquad\epsilon\rightarrow+0.\end{equation}
 Now Eq. (\ref{LinRespStep}) becomes \begin{equation}
\left\langle \sigma_{z}\right\rangle _{t}=\xi_{0}\int_{t}^{\infty}d\tau L(\tau)=\xi_{0}\int_{-\infty}^{\infty}\frac{d\omega}{2\pi}\frac{e^{-i\omega t}}{i\omega+\epsilon}\chi(\omega).\label{sigtviachi}\end{equation}

The imaginary part of the susceptibility is related to the spin-spin
CF by the fluctuation-dissipation relation \begin{equation}
\chi^{\prime\prime}(\omega)=\frac{1}{\hbar}\tanh\frac{\hbar\omega}{k_{B}T}\int_{-\infty}^{\infty}dt\cos(\omega t)S_{zz}(t).\label{FDT}\end{equation}
To calculate the step response, Eq. (\ref{sigtviachi}), one needs
the full susceptibility, and $\chi^{\prime}(\omega)$ can be found
from the Kramers-Kronig relation \begin{equation}
\chi^{\prime}(\omega)=\frac{1}{\pi}\mathcal{P}\int_{-\infty}^{\infty}d\omega^{\prime}\frac{\chi^{\prime\prime}(\omega^{\prime})}{\omega^{\prime}-\omega}.\label{Kramers-Kronig}\end{equation}
Substituting Eq. (\ref{FDT}) into Eq. (\ref{Kramers-Kronig}) one
obtains \begin{equation}
\chi^{\prime}(\omega)=\int_{-\infty}^{\infty}dt\, S_{zz}(t)f(\omega,t),\label{chire}\end{equation}
 where \begin{eqnarray}
f(\omega,t) & = & \frac{1}{\pi\hbar}\mathcal{P}\int_{-\infty}^{\infty}\frac{d\omega^{\prime}}{\omega^{\prime}-\omega}e^{i\omega^{\prime}t}\tanh\frac{\hbar\omega^{\prime}}{k_{B}T}\nonumber \\
 & = & \frac{1}{\pi\hbar}\mathcal{P}\int_{-\infty}^{\infty}\frac{dz}{z}e^{i\left(\omega+z\right)t}\tanh\frac{\hbar\left(\omega+z\right)}{k_{B}T}.\end{eqnarray}
This can be rewritten in the form with no singularity in the integrand
at $z\rightarrow0$:\begin{eqnarray}
f(\omega,t) & = & \frac{1}{\pi\hbar}e^{i\omega t}\int_{0}^{\infty}\frac{dz}{z}\left[e^{izt}\tanh\frac{\hbar\left(\omega+z\right)}{k_{B}T}\right.\nonumber \\
 &  & \qquad-\left.e^{-izt}\tanh\frac{\hbar\left(\omega-z\right)}{k_{B}T}\right].\label{eq:-1}\end{eqnarray}
 At high temperatures one can expand the $\tanh$ terms and after
integration obtain\begin{equation}
f(\omega,t)=\frac{1}{\pi k_{B}T}e^{i\omega t}\left[i\pi\omega\mathrm{sign}(t)+2\pi\delta(t)\right].\label{eq:}\end{equation}
 Inserting this into Eq. (\ref{chire}) and taking into account that
$S_{zz}(t)$ is an even function, one obtains\begin{equation}
\chi^{\prime}(\omega)=\frac{2}{k_{B}T}\left[1-\omega\int_{0}^{\infty}dt\,\sin(\omega t)S_{zz}(t)\right].\label{eq:-2}\end{equation}
Combining this with Eq. (\ref{FDT}) at high temperatures \begin{equation}
\chi^{\prime\prime}(\omega)=\frac{2\omega}{k_{B}T}\int_{0}^{\infty}dt\cos(\omega t)S_{zz}(t)\end{equation}
yields \begin{equation}
\chi(\omega)=\frac{2}{k_{B}T}\left[1+i\omega\int_{0}^{\infty}dt\, e^{i\omega t}S_{zz}(t)\right].\end{equation}

Now for the step response from Eq. (\ref{sigtviachi}) follows \begin{equation}
\left\langle \sigma_{z}\right\rangle _{t}=\frac{2\xi_{0}}{k_{B}T}\int_{-\infty}^{\infty}\frac{d\omega}{2\pi}\frac{e^{-i\omega t}}{i\omega+\epsilon}\left[1+i\omega\int_{0}^{\infty}dt^{\prime}\, e^{i\omega t^{\prime}}S_{zz}(t^{\prime})\right].\end{equation}
 Here the first term vanishes and in the second term one can set $\epsilon=0$.
This yields \begin{equation}
\left\langle \sigma_{z}\right\rangle _{t}=\left\langle \sigma_{z}\right\rangle _{\mathrm{eq}}S_{zz}(t),\end{equation}
where $\left\langle \sigma_{z}\right\rangle _{\mathrm{eq}}=2\xi_{0}/(k_{B}T)$
is the equilibrium spin polarization.

\bibliographystyle{epj}
\bibliography{chu-own,gar-own,gar-relaxation,gar-tunneling,gar-MM-ordering,gar-books}

\end{document}